# REJOINDER OF: TREELETS—AN ADAPTIVE MULTI-SCALE BASIS FOR SPARE UNORDERED DATA

By Ann B. Lee, Boaz Nadler and Larry Wasserman

*Carnegie Mellon University, Weizmann Institute of Science and Carnegie Mellon University*

We are grateful to all of the discussants for their thoughtful comments. Their remarks have added significant insight and perspective on the work. As a variety of issues have been raised, we have organized our rejoinder according to main topics that have been brought up by the discussants.

**1. A multiresolution transform guided by the second-order statistics of the data.** The treelet transform is a multiresolution transform that allows one to represent the original data in an alternative form. Rather than describe the data in terms of the original set of covariates, we perform a series of rotations which gradually reveal the hierarchical grouping structure of the covariates. The idea is very similar to the Grand Tour by Asimov (1985). The treelet transform is a tour "guided" by the covariance structure of the data.

Once the treelet transform has been completed, there are multiple ways of choosing an orthogonal basis (see Section 2.2). We never directly discard residual terms as noise. These terms are in fact an integral part of the final representation. In the simulated example of Section 4.2, most of the detail variables represent noise with small expansion coefficients; consequently, only certain coarse-grained variables are chosen for regression. In general, however, detail variables may convey crucial information. The latter point is illustrated in Sections 5.1 and 5.3, where we use the standard choice of one scaling term and $p-1$ difference terms; that is, an observation $\mathbf{x}$ is decomposed according to

$$\mathbf{x} = s\phi + \sum_{i=1}^{p-1} d_i \psi_i,$$

where the first term is a coarse-grained representation of the signal and the $d$-terms represent "differences" between node representations at two consecutive levels in the tree.

**2. Orthogonal versus overcomplete bases.** Tibshirani and Bickel/Ritov correctly point out that one need not restrict attention to one treelet level. An overcomplete dictionary of treelets can certainly be used for prediction. The "tree harvesting" scheme by Hastie et al. (2001), for example, takes node averages of all $2p-1$ nodes in a hierarchical tree and uses these averages as new predictors for regression. The same scheme could be applied to treelets, but one would then also lose some of the strengths of treelets: Regression/classification is just one application of the treelet transform. More generally, the method yields a multiresolution analysis and a *coordinate system* of the data: we have a multi-scale basis function expansion of the data $\mathbb{X} = (\mathbf{x}_1, \ldots, \mathbf{x}_n)$ and the covariance matrix $S = \frac{1}{n}\mathbb{X}^T\mathbb{X}$. An *orthonormal* basis also has many advantages compared to an overcomplete basis: (i) The representation is easy to interpret and computationally simple, (ii) the solution is *stable* in the sense that adding or omitting a covariate does not change the fit of the other covariates, (iii) the theoretical analysis is much simpler and (iv) the expansion coefficients sometimes carry information on the effective dimension of the data set and the relative importance of the coordinates; removing terms with small coefficients then has the effect of regularizing and denoising the data.

**3. Treelets versus averaging predictors on preclustered trees.** Meinshausen and Bühlmann ask how treelets are different from the following scheme: First order variables in a hierarchical cluster tree (under, e.g., complete linkage) and then find a basis on the tree by Principal Component Analysis (PCA). Tibshirani suggests a related scheme where one first builds a hierarchical cluster tree and then simply averages predictors in each cluster. Tuglus and van der Laan suggest other more sophisticated clustering techniques. We have not completed a full comparison of treelets and the schemes proposed by the discussants but would like to mention a few theoretical and practical advantages of treelets.

First, there are relatively few theoretical results on hierarchical clustering algorithms. Many popular procedures are not stable to noise, or even consistent. In Hartigan (1981), Hartigan writes that "standard hierarchical techniques, such as average and complete linkage, are hopelessly inconsistent [for density estimation]"; he then shows that single-linkage clustering is only weakly consistent or "fractionally consistent." Unfortunately, even less is known about the statistical properties of more complex methods that combine hierarchical clustering, averaging of predictors and regression. The treelet method has the advantage of being simple. The construction of an orthogonal basis and a tree in one step, using the covariance structure of
the data, makes the algorithm computationally efficient and the method
amenable to theoretical analysis. In our paper, we examine the large sample
properties of treelets. We also show, for a block covariance model, that the
required sample size for detecting the underlying population tree structure
is logarithmic rather than linear in $p$. It is not clear if the same results apply
to PCA on pre-clustered trees. It would be interesting to see more theoretical results on the many promising hierarchical clustering algorithms that
have been suggested in the literature.

Compared to "simple averaging" of predictors in clusters, treelets also
have other advantages: (i) The method yields an orthonormal basis; see
item 2 above. (ii) There is information *in the basis functions themselves*.
Simple averaging does not provide such information and can also not adapt
to the data. Treelets are constant on groups of indistinguishable variables
(see Section 3.2.1, Corollary 1); this is not the case for simple averaging
where the loadings are sensitive to the *exact order* in which one merges
the variables. Moreover, if the groupings are less well defined and more
"fuzzy," the loadings in treelets will also adapt accordingly. The latter point
is illustrated by the waveforms in Figures 6, 7 and 10.

**4. Identifiability and uniqueness. Sparse PCA.** Bickel and Ritov (BR)
raise two theoretical issues: identifiability and uniqueness of treelets. As BR
point out, the treelet transform $T(\Sigma)$ viewed as a population parameter
is a function of the population covariance matrix $\Sigma$ only. The underlying
structure in linear mixture models is indeed nonidentifiable, as there exist
more than one solution for the loading vectors. The treelet transform chooses
a representation that reflects groupings of highly correlated variables. These
groups of variables, however, do not have to be disjoint for an approximate
block covariance structure (as Example 3 in the paper also shows).

Why do we need treelets and what is the advantage of a treelet transform compared to other sparse methods? A notorious difficulty of least
squares and variable selection methods lies in the collinearity between covariates; see Fan and Lv (2008), Section 4.1.2, on the need of a transform that
takes advantage of the joint information among the predictors. Sparse PCA
[Zou, Hastie and Tibshirani (2006)] with a combined $l_1$- and $l_2$-penalization
scheme does find groupings of correlated variables but the results depend
on the *particular* choice of tuning parameters. The latter choice defines the
scale of the analysis. Real data sets, however, are often rather complex and
groupings can occur on multiple scales. One of the strengths of the treelet
method is that it captures hierarchical groupings by construction. The series of transformations in the method helps weaken correlation among the
covariates. We do not think that treelet transform is a replacement of other sparse methods. On the other hand, it can be a useful tool if combined with other sparse methods as suggested by Fan and Lv.

Bickel and Ritov also raise the issue of uniqueness. We would like to point out that if we use covariances as a similarity measure, the treelet transform is unique up to a permutation of second-order statistically exchangeable variables. In most applications, correlations seem to be a better measure of similarity. The treelet transform $T$ with a correlation measure is, however, multivalued: formally, $T(P)$ is a set of transforms rather than a single transform. If treelets are viewed as an exploratory tool, then we do not find this fact troubling. An analogy with mixture models might be helpful. Mixture models are famous for suffering numerous irregularities: local nonidentifiability, intractable limiting distributions of test statistics, nonunique maxima of the likelihood function, infinite likelihood values and slow convergence rates, to name a few. For theoretical analysis, they can be a nightmare. Nonetheless, they are used in many applications with great practical success. Like BR, we find the nonidenitifability and multivalued properties of treelets disquieting but, like mixture models, they nonetheless do seem to be useful. Ultimately, the effectiveness of treelets in real problems will determine their utility. On the other hand, any theoretical ideas that provide insight are welcome. Thus, we are intrigued by BR's conjecture at the end of their Section 1. We look forward to hearing about future progress on this idea.

**5. Supervised learning.** We agree with Meinshausen and Bühlmann (MB) that constructing predictors without using the response $Y$ does fail in some cases. The advantage of treelets is the intepretability of the derived features. Sometimes constructing predictors without reference to $Y$ is a necessity. An example is the problem of semi-supervised inference. In this case we observe labeled data $(X_1, Y_1), \ldots, (X_n, Y_n)$ but we also have access to unlabeled data $X_{n+1}, \ldots, X_N$, where $N$ is much larger than $n$. Evidence that the unlabeled data alone can be used to construct effective predictors abounds in the machine learning literature. As Tibshirani writes in his discussion, there is also growing empirical evidence that unsupervised feature extraction can provide an effective set of features for supervised learning. Tibshirani cites the recent work by Hinton, Osindero and Teh (2006) on learning algorithms for Boltzmann machines as an example.

MB point out that information in the response variable can be used in various ways "ranging from weak use of the response to stronger involvement." They give some innovative suggestions on how the response could potentially be incorporated into a treelet framework. As MB writes, the current supervised choice of basis functions by cross-validation represents one use of the response, but perhaps a weaker one. In their discussion, they mention "fully supervised" schemes where $Y$ is used to construct the groupings themselves. We plan to look into various such extensions of treelets in the future.

Regarding supervised learning of predictors, we are intrigued by Bickel and Ritov's suggested method for iteratively growing a class of basis functions. Independently, we have been experimenting with a similar algorithm in the context of modeling phenotypes on interactions of SNPs. Like BR, we start with main effects and gradually add interaction terms in an adaptive fashion. We have recently begun a theoretical analysis of this idea and we look forward to comparing our results with those of BR.

**6. Scalability and other computational issues.** Murtagh raises questions about the scalability of the treelet algorithm. Our current implementation of the treelets uses an exhaustive search at each level of the tree. This is typical of bottom-up hierarchical algorithms and corresponds to a computational cost of $O(Lp^2) + m$, where $L$ is the level of the tree, $p$ is the number of variables (or leaves in the tree) and $m$ is the initial cost of computing the data covariance matrix. However, by keeping track of local changes in the covariance matrix (see Section 2.1), the complexity of the treelet algorithm can further be reduced to $O(Lp) + m$.

We do not believe our method has any computational disadvantage compared to Murtagh's method with fixed Haar wavelets on precomputed dendrograms [Murtagh (2007)]. The cost in computing an adaptive basis is neglible compared to the cost of computing the dendrogram itself. The experimental evaluations in the paper are on $p = 1000$ variables because of the nature of the problems and, in the case of the analysis of the Golub microarray data, because of the availability of benchmark results for this choice of $p$. One can run the computations efficiently in higher dimensions, such as $p \gtrsim 10000$. While we disagree with Murtagh regarding scalability, we agree that treelets may not be appropriate for "ultra-high" dimensional settings (e.g., $p \gtrsim 100000$), where certain topological phenomena may dominate the data.

We plan to post open-source code in both $C++$ and $R$ at http://www.stat.cmu.edu/~annlee/software.htm by the end of the summer of 2008. Until then, we have made available some Matlab test code at the same URL. This code, however, has not been optimized for speed or efficiency in memory use.

**7. Applicability to microarray data.** Finally, Qiu and Tuglus/van der Laan (TV) comment on the applicability of treelets to microarray data. We are not experts on the analysis of such data, but would like to bring up a few potentially important points.

TV correctly state that treelets are built upon a hierarchical scheme of grouping variables and that the graph structure is solely based on correlations. They suggest that other similarity or distance measures may be more appropriate for clustering. We agree on this point but would like to emphasize that the goal of treelets is not clustering per-se. It is the construction of a multi-resolution representation of data. Should other distance measures be used, one needs to define how to aggregate the resulting sets of variables. In principle, one can also think of graph-theoretic measures of similarity between variables, and nonlinear treelet-inspired local transformations between them (for example, for data lying on nonlinear manifolds). The theoretical analysis becomes increasingly difficult once one goes beyond second-order statistics.

Qiu remarks that a possible pitfall of the treelet methods is its preference for sum variables with higher variance than the corresponding detail variables. He argues that genes with smaller variability may be the ones responsible for essential biological functionalities. In our framework, detail variables are not discarded. They are only removed from further merging in the tree. These detail variables can certainly be included in a regression or classification model, as is also shown in the paper. Furthermore, correlation-based treelets can actually be useful in unraveling groups of genes with low variance. Consider, for example, data with sets of genes with very different variances and different intrinsic noise levels. Global variance-based methods such as PCA or sparse PCA would not pick up groups of genes with individual low variances. However, if these variables are highly correlated, they will be among the first ones to be identified and merged with the treelet algorithm.

In our paper (Section 5.3) we describe a "two-way" classification scheme for the Golub leukemia data set. Qiu asks for a clarification of this method. Our main goal here was to show that treelets can be built on both variables (genes) and samples (patients). We are not claiming that the method is superior—only that a general method such as treelets can be competitive with state-of-the-art algorithms that are especially tuned for the analysis of microarray data. The proposed scheme is as follows: First compute treelets on the genes using the training data. This part is the same as for "LDA on treelet features." The second step, however, is different. Here we express *all* 72 samples (patients) in terms of their new profiles over the $K$ maximum variance treelets. We build treelets on the new patient profiles and find the two main branches of the tree. The two groups represent the two cancer classes (ALL or AML); these groups are labeled using the training data and a majority vote. The error is evaluated on the test set (see Figure 9, right). Note that the second step, the labeling of samples, is an example of *semi-supervised learning* (see item 5). It is not a violation of cross-validation. On the contrary, semi-supervised learning (SSL) is a powerful method that is often used to improve classification; see, for example, Belkin and Niyogi (2005). The key idea behind SSL is that *unlabeled* data can be used to uncover the underlying structure of the data (e.g., low-dimensional manifolds, groupings etc.) and that this knowledge can lead to better prediction than if only labeled data had been used.

To summarize, we do not claim that the treelets are the optimal method to model microarray data. They might miss important effects in certain settings. However, treelets or some of their possible generalizations may turn out to be useful in the analysis of such data. Further research is required in this direction.

**Acknowledgments.** It has been a real pleasure to participate in this discussion. We would like to thank the AOAS Biology Sciences Editor Michael Newton, the AOAS Editor-In-Chief Bradley Efron and the discussants for making this exchange of ideas possible.

## REFERENCES

Asimov, D. (1985). The Grand Tour: A tool for viewing multidimensional data. *SIAM J. Sci. Comput.* **6** 128–143.

Belkin, M. and Niyogi, P. (2005). Semi-supervised learning on Riemannian manifolds. *Machine Learning* **56** 209–239.

Fan, J. and Lv, J. (2008). Sure independence screening for ultra-high dimensional feature space. *J. Roy. Statist. Soc. B.* To appear.

Hartigan, J. A. (1981). Consistency of single linkage for high-density clusters. *J. Amer. Statist. Assoc.* **76** 388–394.

Hastie, T., Tibshirani, R., Botstein, D. and Brown, P. (2001). Supervised harvesting of expression trees. *Genome Biology* **2** research0003.1–0003.12.

Hinton, G. E., Osindero, S. and Teh, Y.-W. (2006). A fast learning algorithm for deep belief nets. *Neural Comput.* **18** 1527–1554.

Murtagh, F. (2007). The Haar wavelet transform of a dendrogram. *J. Classification* **24** 3–32.

Zou, H., Hastie, T. and Tibshirani, R. (2006). Sparse principal component analysis. *J. Comput. Graph. Statist.* **15** 265–286.

A. B. Lee
L. Wasserman
Department of Statistics
Carnegie Mellon University
Pittsburgh, Pennsylvania 15206
USA
E-mail: annlee@stat.cmu.edu
  larry@stat.cmu.edu

B. Nadler
Department of Computer Science
  and Applied Mathematics
Weizmann Institute of Science
Rehovot
Israel
E-mail: boaz.nadler@weizmann.ac.il